\documentclass[a4paper]{article} \fontsize{12}{15}
\usepackage[dvipdfm]{graphicx}
\usepackage{wrapfig}

\begin{document}
\title{Positronium Hyperfine Splitting}

\author{Akira Miyazaki,\\
        {\it {\small Department of Physics, Graduate School of Science}}\\
        {\it {\small and International Center for Elementary Particle Physics (ICEPP),}}\\
        {\it {\small The University of Tokyo, 7-3-1 Hongo, Bunkyo-ku, Tokyo, 133-0033, Japan}} \\ 
        {\bf {\small miyazaki@icepp.s.u-tokyo.ac.jp}}} 
\date{}


\maketitle

\begin{abstract}
Positronium is an ideal system for the research of QED in the bound state.
The hyperfine splitting of positronium (Ps-HFS: about 203 GHz) is a good tool to test QED 
and also sensitive to new physics beyond the Standard Model.
Previous experimental results show 3.9\,$\sigma$ (15 ppm) discrepancy from the QED $\mathrm{O}\left(\alpha ^3 \ln{1/\alpha}\right)$ 
 prediction.
We point out probable common systematic errors in all previous experiments.
I measure the Ps-HFS in two different ways.
(1) A prototype run without RF system is described first.
(2) I explain a new direct Ps-HFS measurement without static magnetic field.
The present status of the optimization studies and current design of the
experiment are described.
We are now taking data of a test experiment for the observation of the direct transition.
\end{abstract}

\section{Introduction}

\begin{figure}[ht]
\begin{minipage}{0.3\hsize}
 \begin{center}
  \includegraphics[width=2cm]{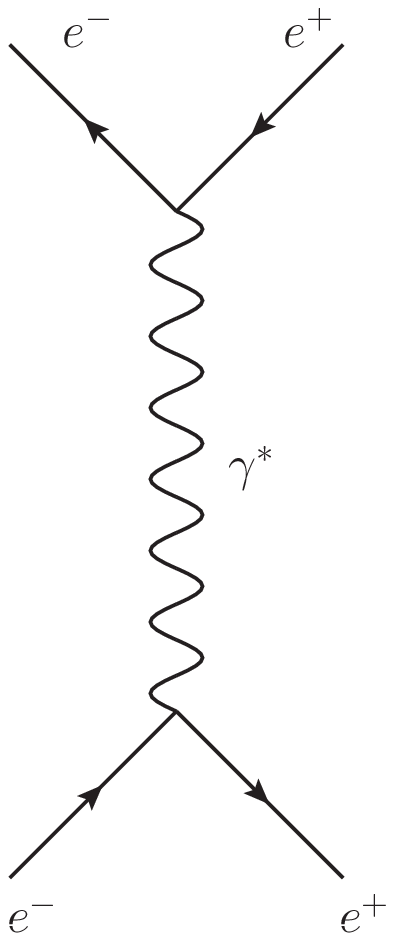}
  \caption{\label{fig:osci} $o$-Ps contribution to Ps-HFS} 
 \end{center}
\end{minipage}
\begin{minipage}{0.7\hsize}
 \begin{center}
  \includegraphics[width=8cm]{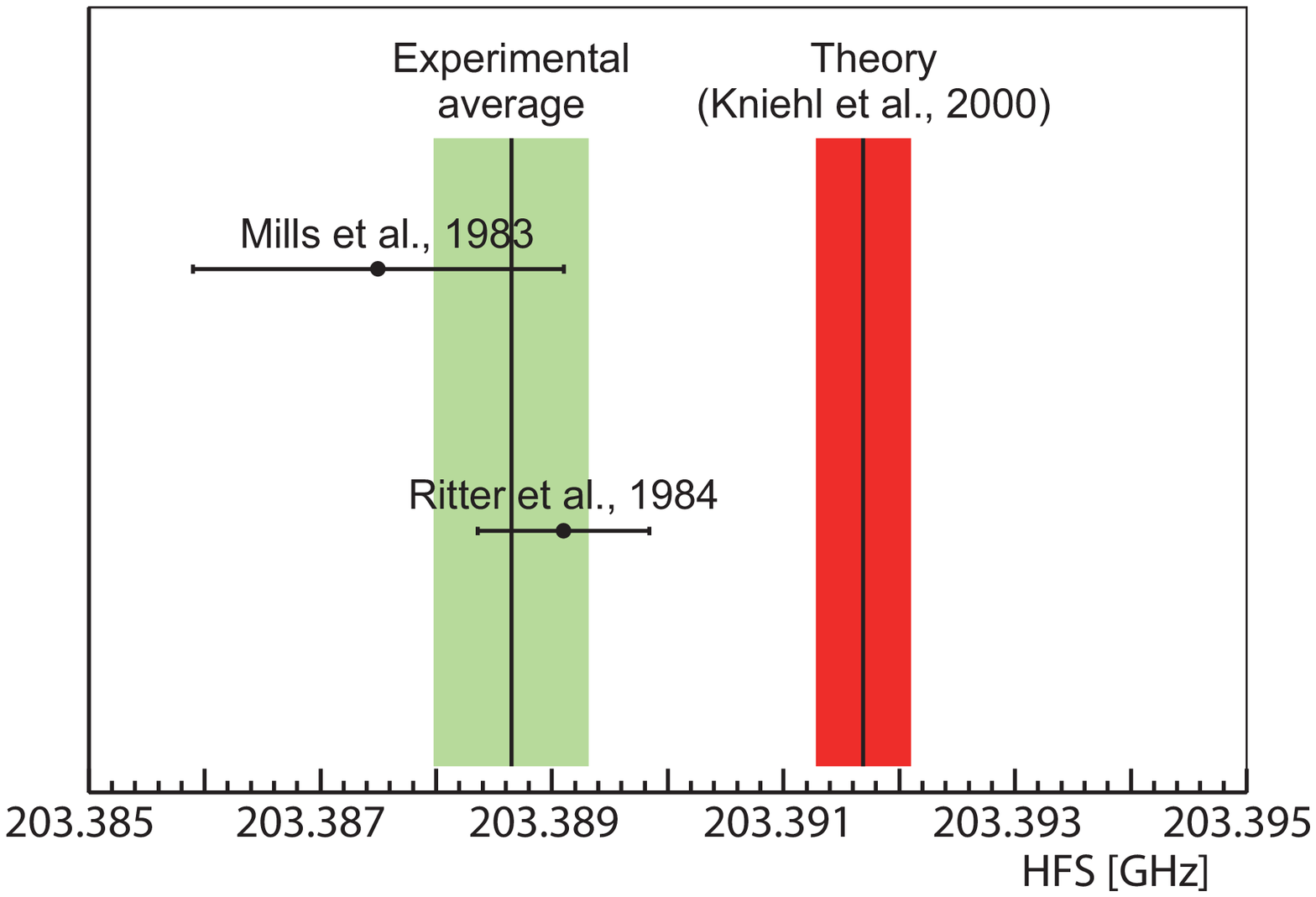}
  \caption{\label{fig:hfs} The discrepancy of Ps-HFS value}
 \end{center}
\end{minipage}
\end{figure}

Positronium (Ps), the electron-positron bound state, is a purely leptonic system.
The energy difference between $ortho$-positronium ($o$-Ps, $^3S_1$ state) and
$para$-positronium ($p$-Ps, $^1S_0$ state)
\footnote{Although $p$-Ps decays mainly into two photons with lifetime of 125~ps,
it takes 142~ns for $o$-Ps to decay. 
This is because $o$-Ps can only decay into three photons which is strongly suppressed by invariant matrix and kinematics.
Two photon decay of $o$-Ps is forbidden by C conservation.}
 is called hyperfine splitting of positronium (Ps-HFS).
It is a good target to study bound state QED precisely.
The Ps-HFS value is approximately 203~GHz (0.84~meV), which is significantly larger than hydrogen HFS (1.4~GHz).
About one third of this large value is contributed by a quantum oscillation as shown in Fig.~\ref{fig:osci}:
$o$-Ps $\rightarrow \gamma^\ast \rightarrow$ $o$-Ps
\footnote{$Ortho$-Ps has the same quantum number as a photon.}
.
Since some hypothetical particles, such as a milicharged particle,
can participate in the quantum oscillation to shift Ps-HFS value,
its precise measurement provides a probe into new physics beyond the Standard Model.

Measurements of the Ps-HFS have been performed in 70's and 80's~\cite{Mills, Ritter}.
The results were consistent with each other, and combined precision of 3.3~ppm is obtained.
They were consistent with $\mathrm{O}\left(\alpha ^2\right)$ calculation of the QED available at that time.
The corrections of $\mathrm{O}\left(\alpha ^3 \ln{1/\alpha}\right)$ 
have been calculated using NonRelativistic QED (NRQED) in 2000~\cite{Kniehl}.
The new prediction is $\Delta^{\rm th}_{\rm HFS} = $203.391~69(41)~GHz,
where the uncertainty is the unknown nonlogarithmic $\mathrm{O}\left(\alpha ^3\right)$ term 
estimated in an analogous way to the HFS of muonium.
This calculated value differs from the measured value of $\Delta^{\rm exp}_{\rm HFS} = $203.388~65(67)~GHz by 3.9$\sigma$ as shown in Fig.~\ref{fig:hfs}.
This discrepancy may be due to common systematic errors in the previous measurements 
or to new physics beyond the Standard Model.

In all previous measurements, the Ps-HFS value was not directly measured, 
since 203~GHz was too high to produce and control.
Zeeman splitting of $o$-Ps has been measured instead.
A static magnetic field makes Zeeman mixing between $m_z=0$ spin state of $o$-Ps and $p$-Ps.
As a result, the energy level of $m_z=0$ state of $o$-Ps becomes higher than $m_z = \pm1$ state.
This Zeeman splitting, which is approximately proportional to Ps-HFS,
is a few GHz frequency under about 1~Tesla magnetic field.
Static magnetic field is applied in RF cavities where positronium is produced. 
Zeeman transition from $o$-Ps of $m_z=\pm1$ to $o$-Ps of $m_z=0$ has been observed.

We point out the following three possibilities as the common systematic errors in these indirect measurements.
\begin{enumerate}
\item They may underestimate the non-uniformity of the magnetic field.
\item The unthermalized $o$-Ps contribution can result in an underestimation of the material effect.
\item RF systems to cause the transition might not be sufficiently stable.
\end{enumerate}

Direct measurement of Ps-HFS without any magnetic fields is a main topic.
I show the status of a prototype experiment in section 3.
It is completely free from the systematic errors of the magnetic field.
A proposal about the second possibility is summarized by A.~Ishida, {\it et al.}~\cite{ishida}.
The third point is also promising.
The experiment with quantum oscillation instead of the RF source is performed.
I report the result of this experiment in section 2.

\clearpage
\section{Measurement with quantum oscillation}
\begin{wrapfigure}{r}{8cm}
 \begin{center}
  \includegraphics[width=5cm]{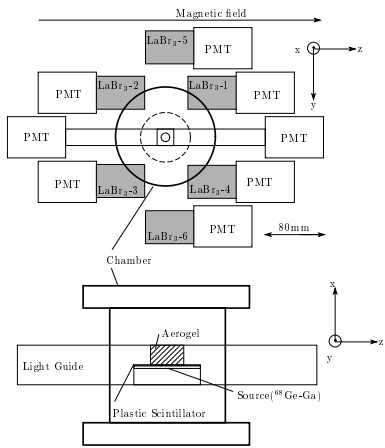}
  \caption{Schematic diagram of the setup\label{fig:setup}}
 \end{center}
\end{wrapfigure}

We used a completely different method from previous one (i.e. without a RF source).
This method is theoretically proposed by V.G.~Baryshevsky~et~al.\cite{Baryshevsky},
in which a quantum oscillation between two Zeeman energy levels of $o$-Ps
is measured in a static magnetic field.
Positrons emitted from a $\beta^+$ source 
are polarized in the direction of their momentum
due to parity violation in the weak interaction.
Consequently, the resulting $o$-Ps is also highly polarized.
This $o$-Ps is a superimposed state of above two energy eigenstates of $o$-Ps in a static magnetic field.
When a perpendicular magnetic field of about 0.1~T is induced to the polarization axis of this $o$-Ps,
this quantum oscillation can be detected as an observable {\it oscillation} in timing spectra of $o$-Ps decay. 

This approach is quite different from the previous all experiments.
It is free from systematic errors originating from the high power light source and the RF cavity with high Q-value.
Instead of them, precise measuring technique of decay curve is crucial.
Especially, time-to-digital converter~(TDC) with high performance 
is essential for precise measurement of the timing spectrum.
Therefore, these two different approaches are complementary.
Both experimental approaches are necessary to understand the discrepancy.

Figures \ref{fig:setup} show a schematic diagram of the experimental setup of quantum oscillation.
The upper figure shows the entire experimental setup.
The magnetic field direction is along the $z$-axis.
The LaBr$_3$(Ce) scintillators are placed in the $yz$-plane.
They detect $\gamma$-rays with high energy resolution of 4.0~\%~(FWHM)  at 511~keV 
and high timing resolution of 200~ps~(FWHM).
The direction of the $\beta^+$ emitted
from the  $^{68}$Ge-Ga source is along the $x$-axis.
The bold circle is a vacuum chamber.
The coordinate system is also shown.
The lower figure is a magnified view of the vacuum chamber, in which the
$^{68}$Ge-Ga source, the thin plastic scintillator and the silica aerogel are located.
This plastic scintillator tags positrons, which go into the silica aerogel.
Positronium is formed in this aerogel to decay into $\gamma$-rays.
The signal from plastic scintillator comes at approximately the formation time of positronium.
The time of the signal from LaBr$_3$ is the decay time.
We took a delayed coinsidence between them
to make a decay curve of positornium.
The time is measured with direct clock TDCs~(5~GHz:~timing resolution of 200~psec).
These TDCs have excellent integral and differential linearities.

\begin{figure}[ht]
\begin{minipage}{0.5\hsize}
\begin{center}
\includegraphics[width=65mm]{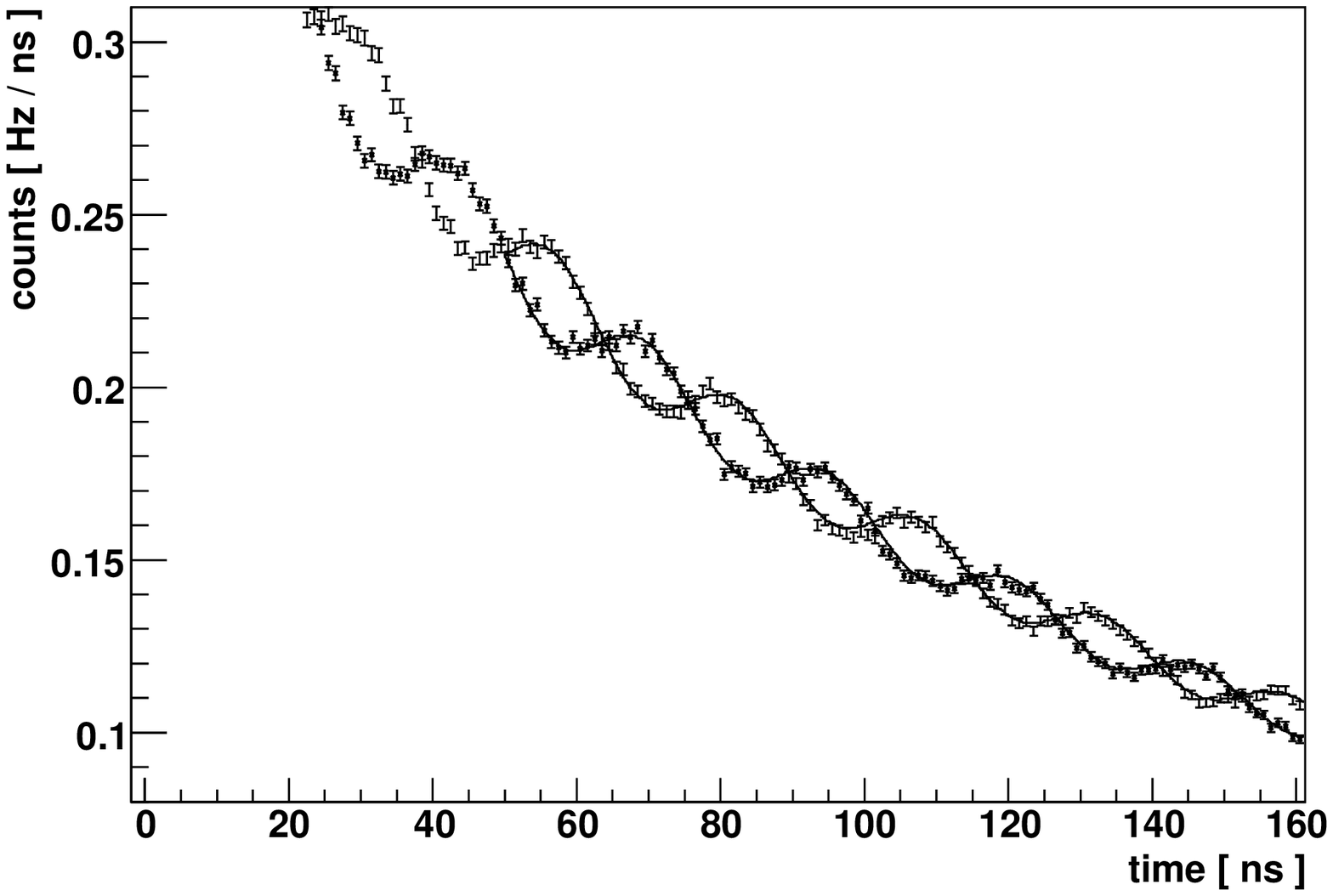}
\end{center}
\end{minipage}
\begin{minipage}{0.5\hsize}
\begin{center}
\includegraphics[width=65mm]{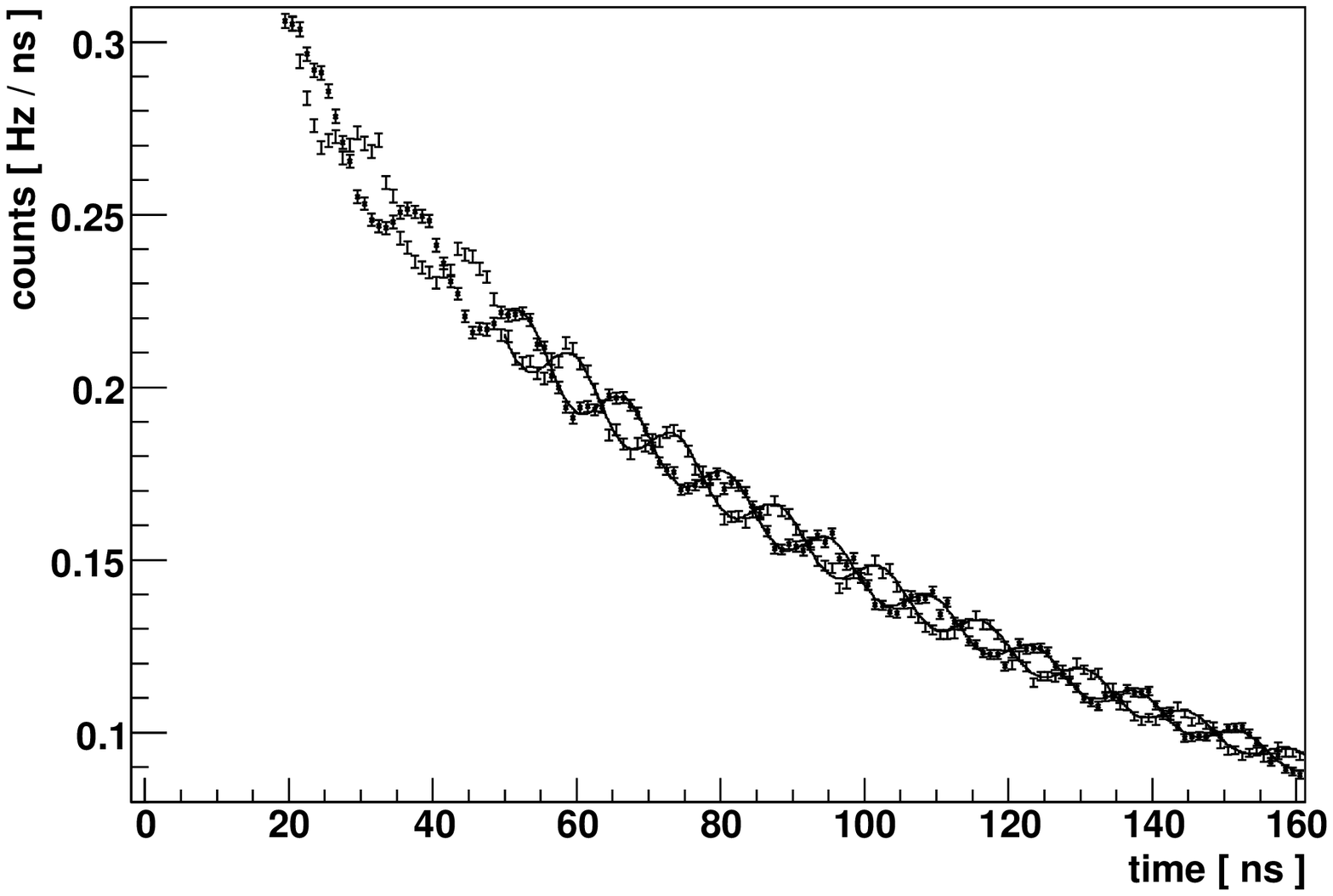}
\end{center}
\end{minipage}
\caption{Timng spectra of the decay curve of $o$-Ps \label{fig:normalfit}}
\end{figure}

Figures \ref{fig:normalfit} show the measured time spectra at a  magnetic value of 100~mT~(left figure) and 135~mT~(right figure).  
In both figures, the data points are plotted with error bars while the solid lines show the best fit results.  
A result of a prototype experiment is obtained as $\Delta^{\rm exp}_{\rm HFS}=$203.324$\pm$0.039(stat.)$\pm$0.015(sys.)~GHz.
The accuracy is 200~ppm, 
which is an improvement by a factor of 90 over the previous experiment 
which used the similar method~\cite{fan}.
This result is consistent with both theoretical calculations and previous precision measurements of transition.
However, we showed that we can improve this result to compete the most precise measurement
with some simple improvements of our detection system.
(in this paper~\cite{uneune})

\section{Direct measurement of Ps-HFS}
We are planing to directly measure the HFS transition,
which does not need a static magnetic field.
It is thus free from the systematic error from the field mentioned in section 1.
However, the direct transition from $o$-Ps to $p$-Ps has very small probability of $3\times10^{-8}$~s$^{-1}$,
since this transition is {\it M1} transition,
and the Ps-HFs is extremely large.
Therefore, a powerful radiational field of 203~GHz is essential so as to stimulate the direct transition.

The frequency of 203~GHz is just intermediate between optical light and radiowave.
There was no high power light source for spectroscopy in sub-THz region.
We are developing a new light system in this region.
A frequency tunable radiational source is necessary to measure a whole shape of the resonance curve
\footnote{RF system in previous precise indirect measurements of 3~GHz was not tunable, neither. 
They changed the strength of a magnetic field to shift the Zeeman splitting
to tune effectively the resonance frequency of Zeeman transition.}
.
Our first target is to just observe the direct transition from $o$-Ps to $p$-Ps
\footnote{This can be detected as an increase of two photon-decay ratio.}
.
In order to accomplish this goal, we are developing following three new {\it optical} devices.
\begin{enumerate}
\item Sub-THz to THz light source called {\it gyrotron},
\item Efficient transportation system of {\it mode converter},
\item Parallel etalon with high-finesse called {\it Fabry-P\'erot cavity}.  
\end{enumerate}
These are explained from the next subsection.

\subsection{Gyrotron}

\begin{figure}[h]
\begin{minipage}{0.5\hsize}
\begin{center}
\includegraphics[width=5cm, angle=-90]{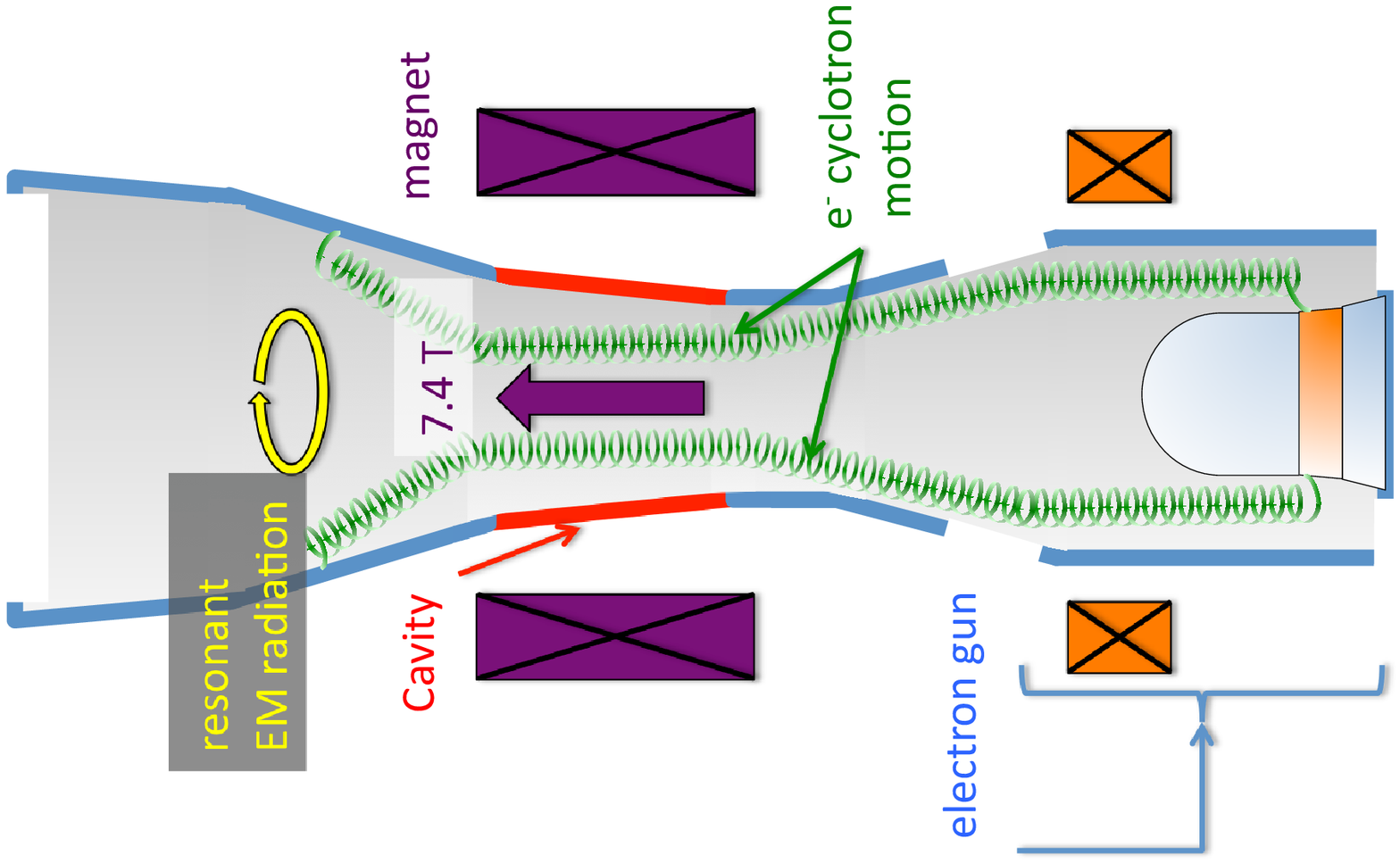}
\caption{\label{gyschematic}Schematic of gyrotron}
\end{center}
\end{minipage}
\begin{minipage}{0.5\hsize}
\begin{center}
\includegraphics[width=5cm]{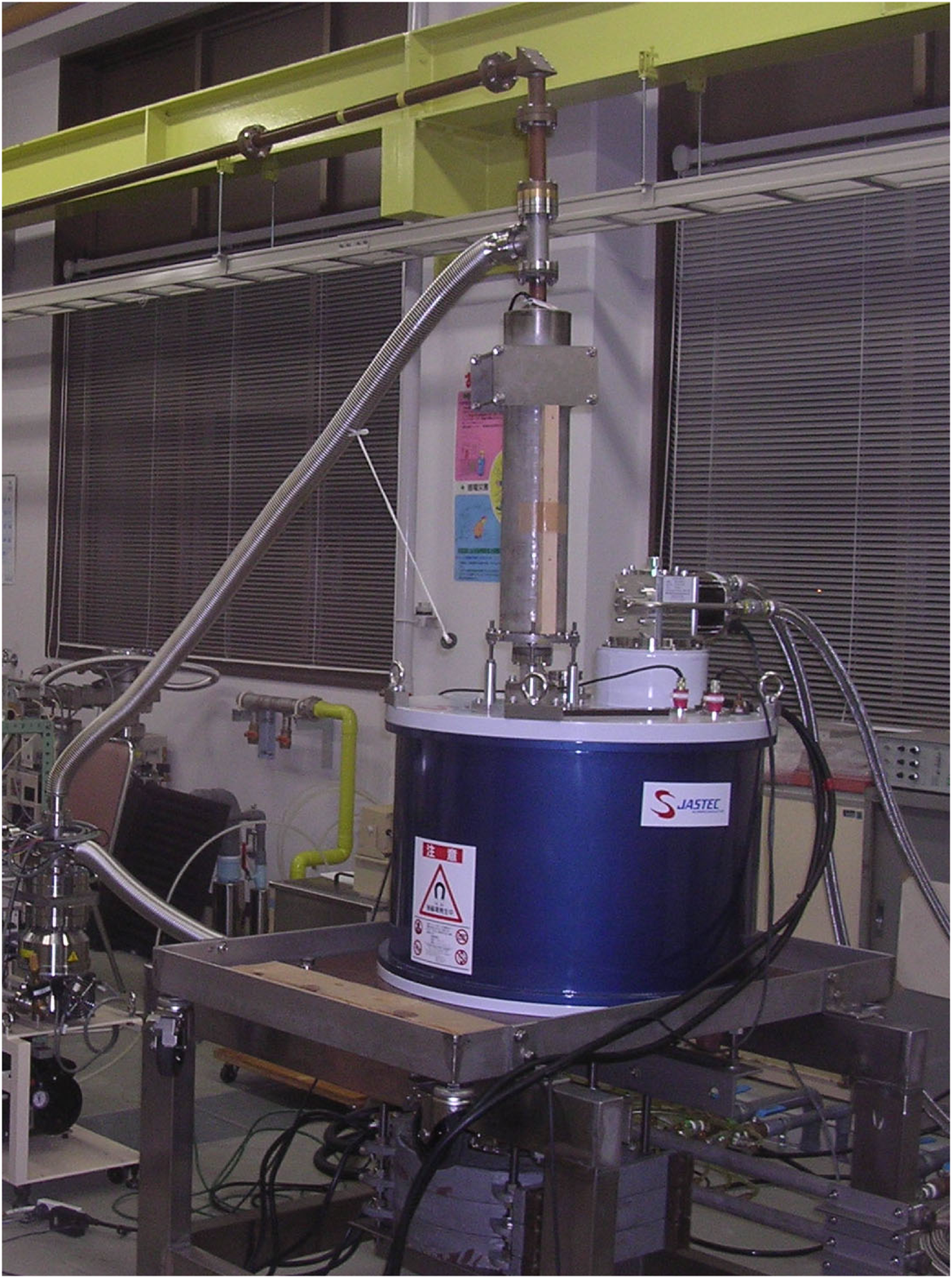}
\caption{\label{gypicture}Picture of the gyrotron for the HFS experiment.}
\end{center}
\end{minipage} 
\end{figure}

\begin{wrapfigure}{r}{4.5cm}
 \begin{center}
  \includegraphics[width=4.0cm, angle=-90]{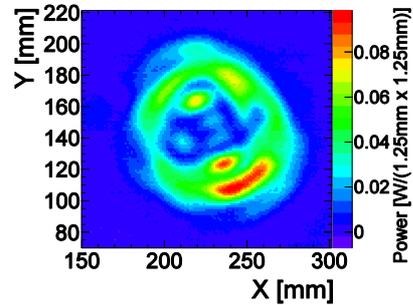}
  \caption{\label{fig:profile}Power profile}
 \end{center}
\end{wrapfigure}

The gyrotron\cite{Idehara} is a novel high power light source for sub-THz to THz frequency region.
It has been developed in the field of nuclear fusion.
The structure of gyrotrons is shown in Fig.~\ref{gyschematic}.
The electrons are emitted from the DC electron gun, concentrated and rotated as cyclotron
motion in the superconducting magnet. The cyclotron frequency $f_c$ is given by
\begin{equation}
	f_c = \frac{eB}{2\pi{}m_0\gamma},
\end{equation}
where $B$ is the magnetic field strength, $m_0$ is the electron rest mass, 
and $\gamma$ is the relativistic factor of the electron.
A cavity is placed at the maximum magnetic field in which resonant frequency is given by 
\begin{equation}
        f =\frac{1}{2\pi}\sqrt{\left(\frac{\chi_{mn}}{R}\right)^2 + \left(\frac{l\pi}{L}\right)^2},
\end{equation}
where $R, L$ are radius and length of the cavity, respectively.
$l$ is an index of longitudinal mode.
$m, n$ are indices of transverse mode.
$\chi_{mn}$ is a root of differential bessel function.
This cavity frequency is tuned just to the cyclotron frequency to enhance the monochromatic light.
The electrons stimulate cyclotron resonance maser in the cavity.
The produced coherent photons are guided to the output port through the window, 
while electrons are dumped at a collector.

We developed a gyrotron operating at $f_c = 203$ GHz with $B=7.364$~Tesla, $\gamma \sim 1.02$, which is shown in Fig.~\ref{gypicture}.
The stable power of 300~W is obtained at the output window of gyrotron.
The frequency width, which is determined by $B$ uniformity and $\gamma$ spread by thermal distribution of
electrons, is expected to be less than 1~MHz. 
It is narrow enough to control the resonance at the Fabry-P\'erot cavity. 
Measured result with a similar gyrotron shows the frequency width is less than 10 kHz~\cite{gywidth}.
Although the frequency can be tuned by changing the $\gamma$ factor with different acceleration of electrons,
the tuning range is limited by the resonant width of the cavity to several hundreds of megahertz.

Figure \ref{fig:profile} shows the power profile of the radiation at taken with an infrared (IR) camera.
The profile has a circular polarization called TE03 mode.
Unfortunately, the mode inside Fabry-P\'erot cavity is a linearly polarized {\it gaussian mode}.
Therefore, the original gyrotron output cannot couple with Fabry-P\'erot cavity.
That's why a mode conversion is necessary to use gyrotron power efficiently.

\subsection{Mode converter}
\begin{figure}[ht]
\begin{minipage}{0.4\hsize}
\includegraphics[width=5cm, angle=-90]{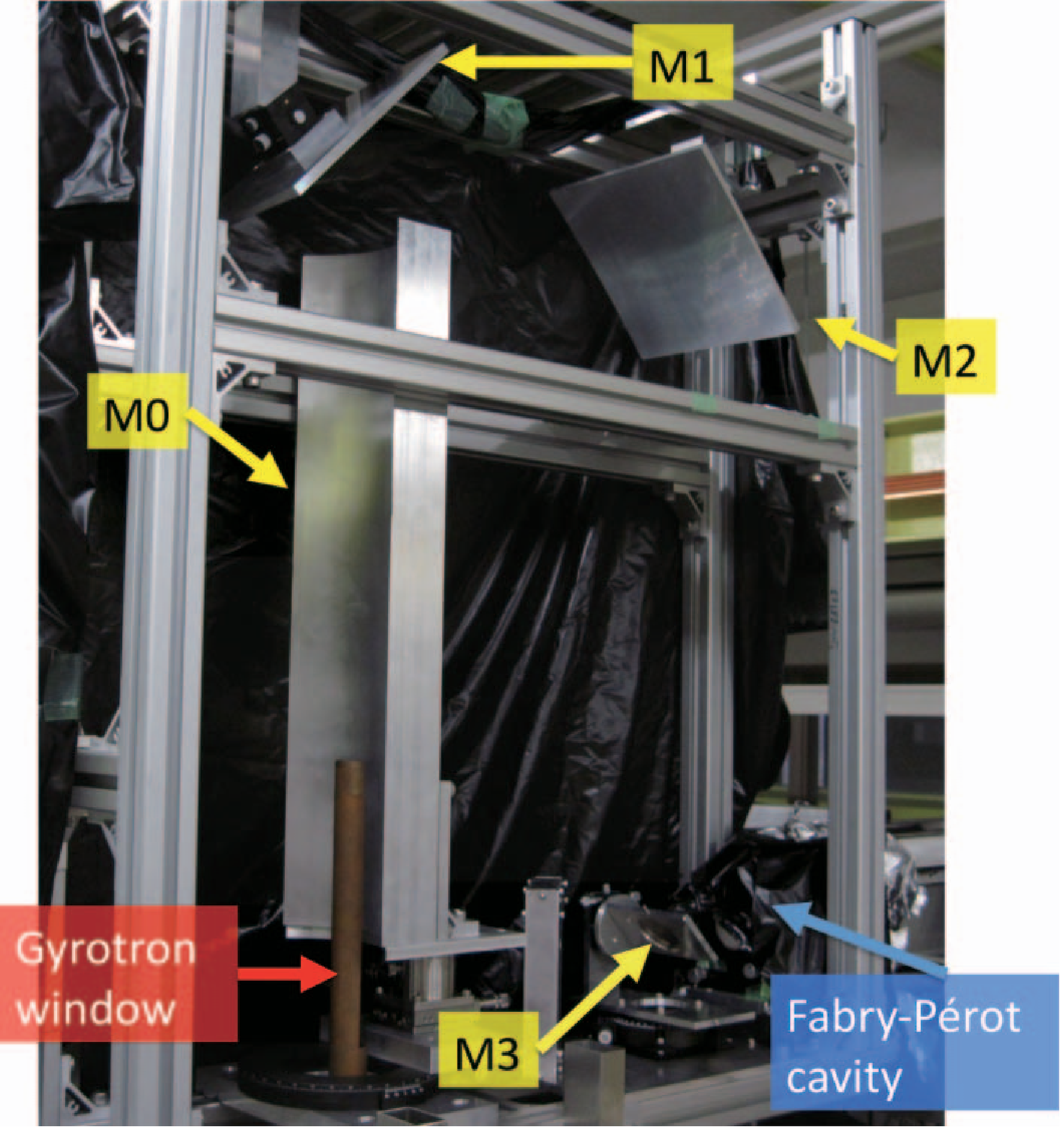}
\caption{\label{fig:trans}Transportation}
\end{minipage}
\begin{minipage}{0.6\hsize}
\includegraphics[clip, width=6cm, angle=-90]{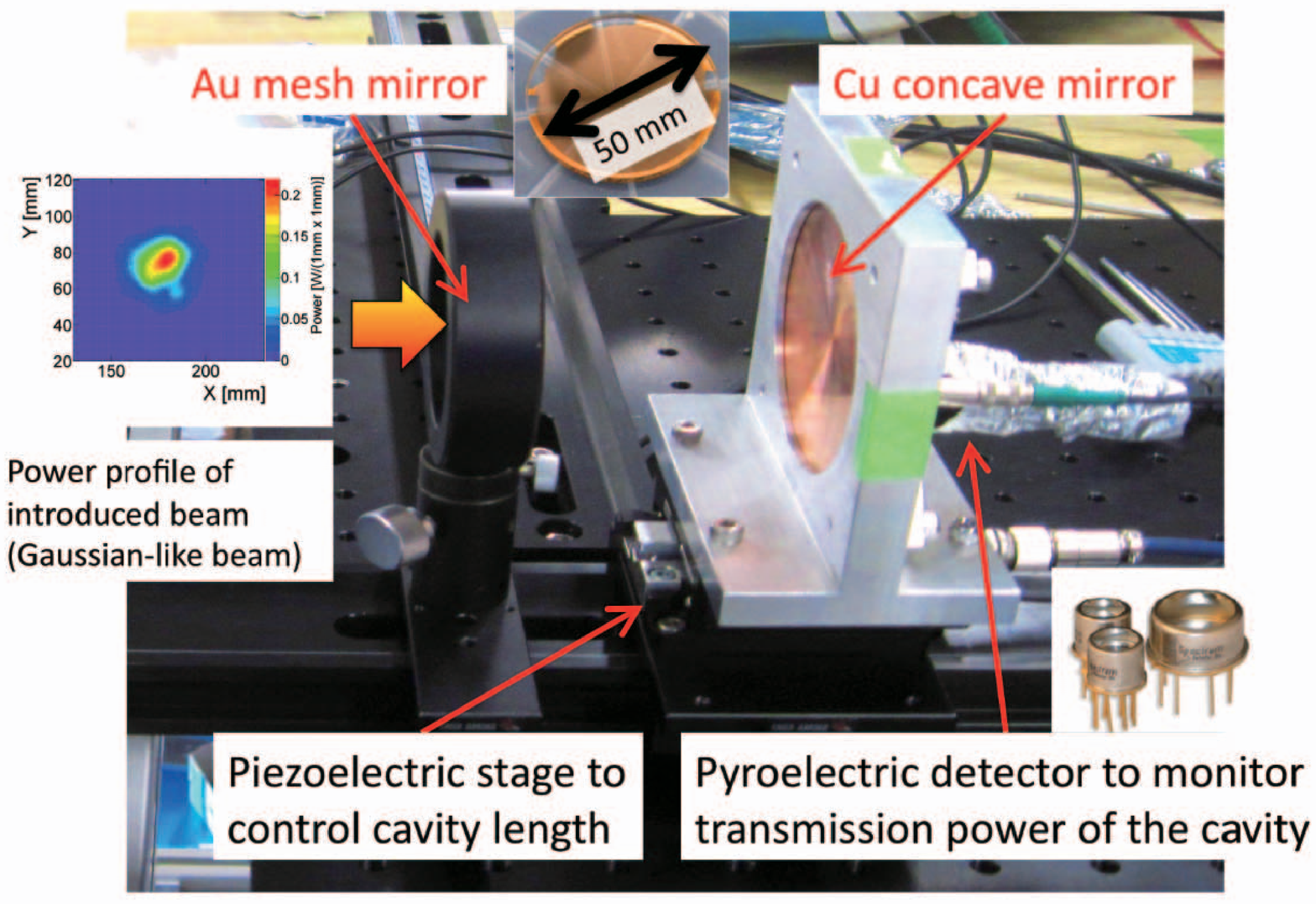}
\caption{\label{fabryphoto}Fabry-P\'erot cavity}
\end{minipage}
\end{figure}

Transportation system is composed of three parabolic mirrors called M0, M1 and M2 as shown in Fig.~\ref{fig:trans}.
The first parabolic mirror M0 converts polarization from circular to linear. 
M1 and M2 simply change the shape of power distribution from bi-gaussian to gaussian. 
Then, plain mirror M3 reflects radiaion and introduces it into Fabry-P\'erot cavity.
The power distribution is successfully converted into gaussian-like mode. 
Coupling between input light and Fabry-P\'erot cavity is now about 60\%.
However, the transformation efficiency is 30\%, 
because the output of the current gyrotron is not optimized.
As a result, about 20\% of the original radiation from gyrotron can resonate in the cavity.

\subsection{Fabry-P\'erot cavity}

Photons produced at the gyrotron are transported and accumulated in a cavity. 
Since 203~GHz photons can be treated optically at the centimeter or larger size scale, 
we use a Fabry-P\'erot cavity, 
which consists of two opposing mirrors to confine photons between them.
Unlike RF cavities, the confinement in the Fabry-P\'erot cavity is 1-dimensional
while the other four sides are open as shown in Fig.\ref{fabryphoto}.
A golden mesh mirror is used on the input side of the cavity to introduce photons from gyrotron.
A copper concave mirror is used on the other side.

The two most important characteristics of a cavity are
{\it finesse} $\mathcal{F}$ and {\it input coupling} C.
With the reflectivity of mesh mirror $R_f$ and concave mirror $R_e$, finesse is defined as
\begin{equation}
 \mathcal{F} = \frac{\pi\left(R_f R_e(1-A)\right)^{1/4}}{1-\left(R_f R_e (1-A)\right)^{1/2}},
\end{equation}
where $A$ is the medium loss inside the cavity.
Round-trip times $N$ of photon in the cavity is given by
\begin{equation}
N = \mathcal{F}/2\pi,
\label{eq:round}
\end{equation}
Therefore finesse characterizes the capability of the cavity to store photons inside.
To maximize the $\mathcal{F}$, power losses must be minimized.
There are 3 types of loss, 
diffraction loss, medium loss and ohmic loss.
With the confinement of photons by the concave mirror,
diffraction loss is negligible in our cavity.
Medium loss in gas\footnote{Mixture of nitrogen 0.9~atm and isobutane 0.1~atm.}
is measured as about 0.1\%.
Ohmic loss occurs at the mirrors, which is around 0.15\% at the
copper mirror and less than 1.0\% at the mesh mirror.

Input coupling is the fraction of input power matched to the cavity mode.
It is an important parameter to efficiently introduce photons into the cavity.
In our cavity, the input coupling is determined by
transmittance of the input mesh mirror.

\begin{figure}[ht]
\begin{tabular}{cc}
\begin{minipage}{0.5\hsize}
\hspace{5pc}
\begin{center}
\includegraphics[width=3.5cm, angle=-90]{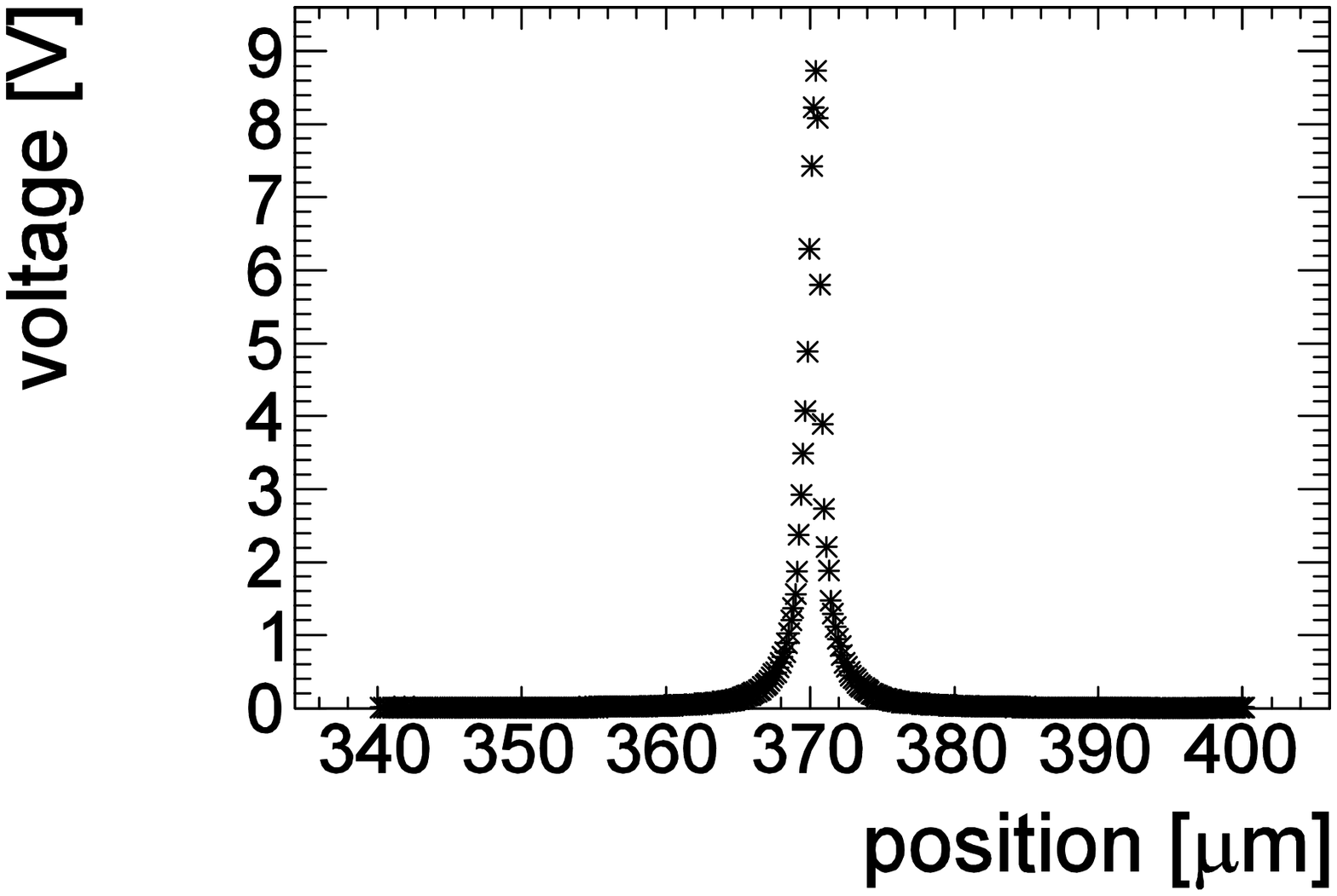}
\caption{\label{fig:touka}Transmission}
\end{center}
\end{minipage} 
\begin{minipage}{0.5\hsize}
\hspace{5pc}
\begin{center}
\includegraphics[width=3.5cm, angle=-90]{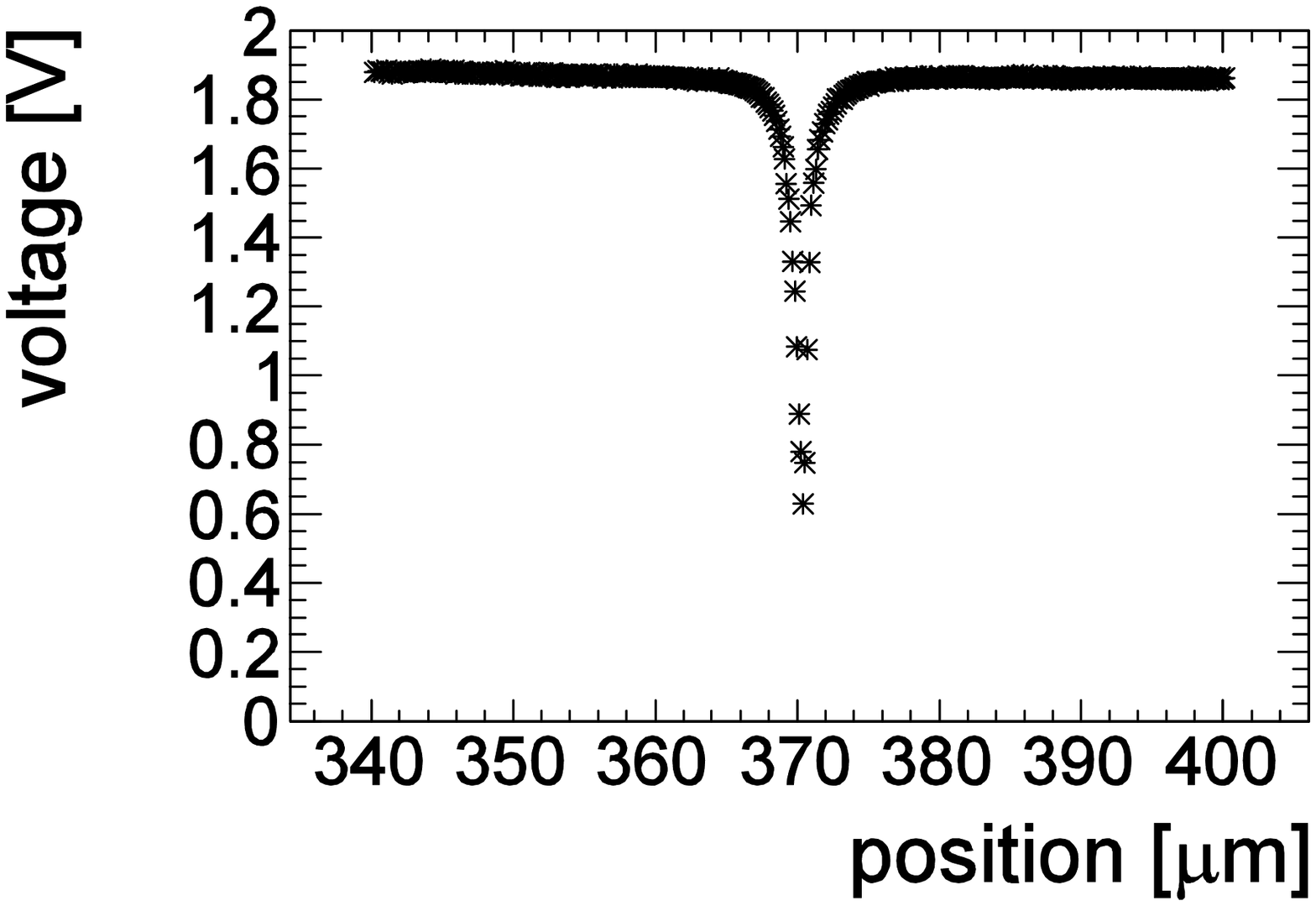}
\caption{\label{fig:hansha}Reflection}
\end{center}
\end{minipage}
\end{tabular}
\end{figure}

Figure \ref{fabryphoto} shows the test setup, 
which is comprised of a mesh mirror on the mirror mount and a concave mirror on a piezo stage.
Transmitted power is measured through a small hole on the concave mirror.
On the other hand, reflected power is measured outside of the mesh mirror.
When we shifted the cavity length precisely by the piezo stage,
Breit-Wigner resonance was observed in the transmitted power monitor as shown in Fig.~\ref{fig:touka}.
Here, the horizontal axis is a position of the piezo stage,
and the vertical axis is the output of power monitor.

Finesse is obtained from the width of the resonance $\Gamma$. Finesse is calculated by
\begin{equation}
 \mathcal{F} = \frac{\lambda/2}{\Gamma},
\end{equation}
where $\lambda=1.47$~mm is wavelegth of 203~GHz.
We got finesse of about $650$, which is equivalent to 100 times round-trip according to Eq.\ref{eq:round}

Figure \ref{fig:hansha} shows the measured reflection power.
Input coupling is given by
\begin{equation}
 C = 1 - \frac{V_{peak}}{V_{baseline}}.
\end{equation}
Here, $V_{peak}$ is a voltage at peak decreasing from $V_{baseline}$, 
the voltage of the baseline of the reflected power.
We achieved input coupling of 67\%.
This large value is mainly due to a good mode conversion explained in the last subsection. 
The current status of the power is summarized in Table \ref{powersum}.
The power of 10~kW is accumulated in the cavity.
\begin{table}[h]
\caption{\label{powersum}
The summary of radiation power with our devices
}
\begin{center}
\begin{tabular}{rrr}
\hline
Device			&	Efficiency	&	power (W)\\
\hline
Gyrotron		&	1		&	300	\\
Mode converter		& 	0.30		&  	90	\\
Fabry-P\'erot cavity	&  0.60 $\times$ 100 $\times$ 2	&	about 10,000\\
\hline
\end{tabular}
\end{center}
\end{table}

\subsection{Detection System}

\begin{figure}[h]
\begin{center}
\includegraphics[height=12cm, angle=-90]{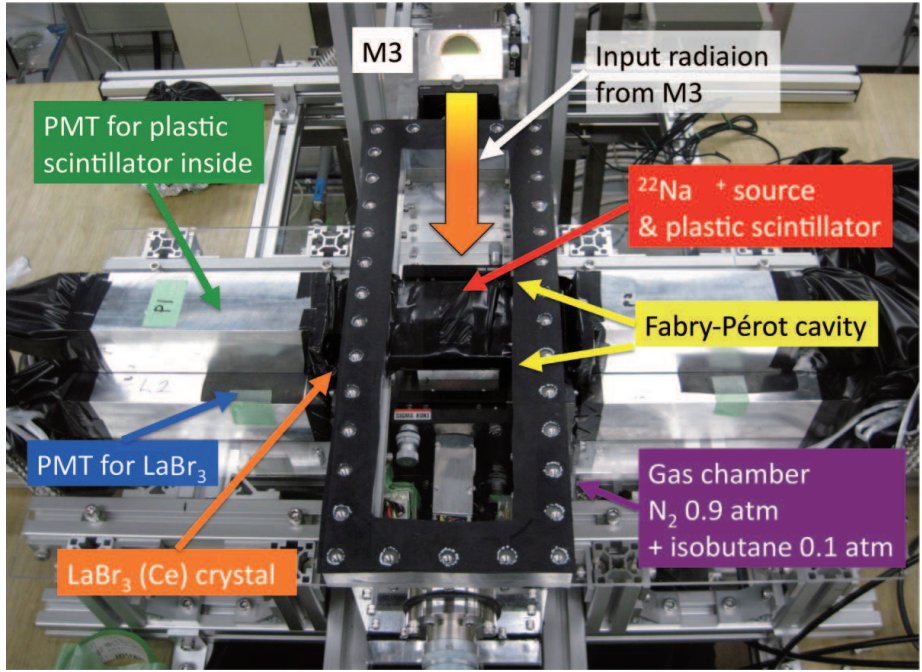}
\caption{\label{fig:detection}A photograph of gas chamber and detection system}
\end{center}
\end{figure}

Figure \ref{fig:detection} shows a photograph of positronium production and signal detection system.
Gyrotron power is introduced to the cavity via the mesh mirror, and accumulated inside Fabry-P\'erot cavity.
This cavity is placed inside a gas chamber filled with mixture gas of 0.9~atm nitrogen and 0.1~atm isobutane.
The positron is emitted from $\beta^{+}$-source.
The $^{22}$Na $\beta^+$-source is located 20~mm above the cavity.
In order to generate start timing, the emitted positrons pass through a $\beta$-tag scintillator, with thickness of 100$\mu$m.
The signal from plastic scintillator transits through light guide made of acryl, to reach photomultiplier (PMT).
Such kind of $\beta$-tagging system was also used in the experiment with quantum oscillation explained in section 2.

A lead collimator, with thickness of 10~mm, is placed under the plastic scintillator
so as to select the positrons which go into the cavity.
It also works as a shield to protect LaBr$_3$(Ce) scintillators from accidental photons
\footnote{They are mainly 1275~keV and 511~keV photons emitted around the source.}
.

Positron forms positronium with an electron in the gas.
$Para$-positronium annihilates into two 511~keV photons immediately,
while $o$-Ps remains with lifetime of $142$ ns to decay into three photons,
whose energy are continuous and less than 511 keV.
A signal of the transition from $o$-Ps to $p$-Ps under 203~GHz is a $delayed$-$two$-$photon$ event.
Four LaBr$_3$ scintillators surround the chamber to detect photons.
Two photon-decay can be easily separated from three photon-decay with this energy information.
The LaBr$_3$ scintillators also have good timing resolution to separate delayed events (i.e. signal of transition)
from prompt events
\footnote{
Almost all the prompt events are two photon-decay.
}
. to improve signal to background ratio significantly
\footnote{
In case of the power of 10~kW in the cavity, 
S/N is estimated to be improved 16 times,
when a timing window is imposed from 50 to 250~ns in decay curve.
}
.

The signal collection efficiency and background rates were estimated using Monte Carlo simulation (GEANT4).
There are three major background processes.
The first one is an three $\gamma$ contamination from $o$-Ps.
The second one is a pick-off background.
A positron in $o$-Ps interacts with a electron in a matter
\footnote{In this case nitrogen and isobutane.}
only to annihilate into two photons.
This process is called pick-off annihilation, and becomes background in our measurement.
The last one is an accidental pileup process.
In order to eliminate these backgrounds, 
we selected back-to-back signal in LaBr$_3$ scintillator
and imposed condition that smeared energy deposit is 511keV$\pm3\sigma$.

The obtained power of 10~kW is used for the simulation.
Figure 13 shows an expected spectra for one month of data taking.
The estimated rate is also summarized in Table 2.
In this table, ''ON'' means the signal under 203~GHz radiation while ''OFF'' means that without radiation
\footnote{In Table 2, total (ON) - total (OFF) is not equal to signal. 
Because, the background events associated with $o$-Ps decreases under high power resonance radiation (i.e. ON).
A Part of $o$-Ps transits into $p$-Ps to decay earlier than the timing window.
As a result, total (ON) - total (OFF) becomes less than expected signal rate.
}
.
The main background is three $\gamma$ contamination.
And the other two backgrounds are the same size of the signal.
We can clearly see the transition within one month.

\begin{tabular}{cc}
&Table 2: The summary of signal estimation\\
\begin{minipage}{0.3\hsize}
\begin{center}
\includegraphics[width=4cm, angle=-90]{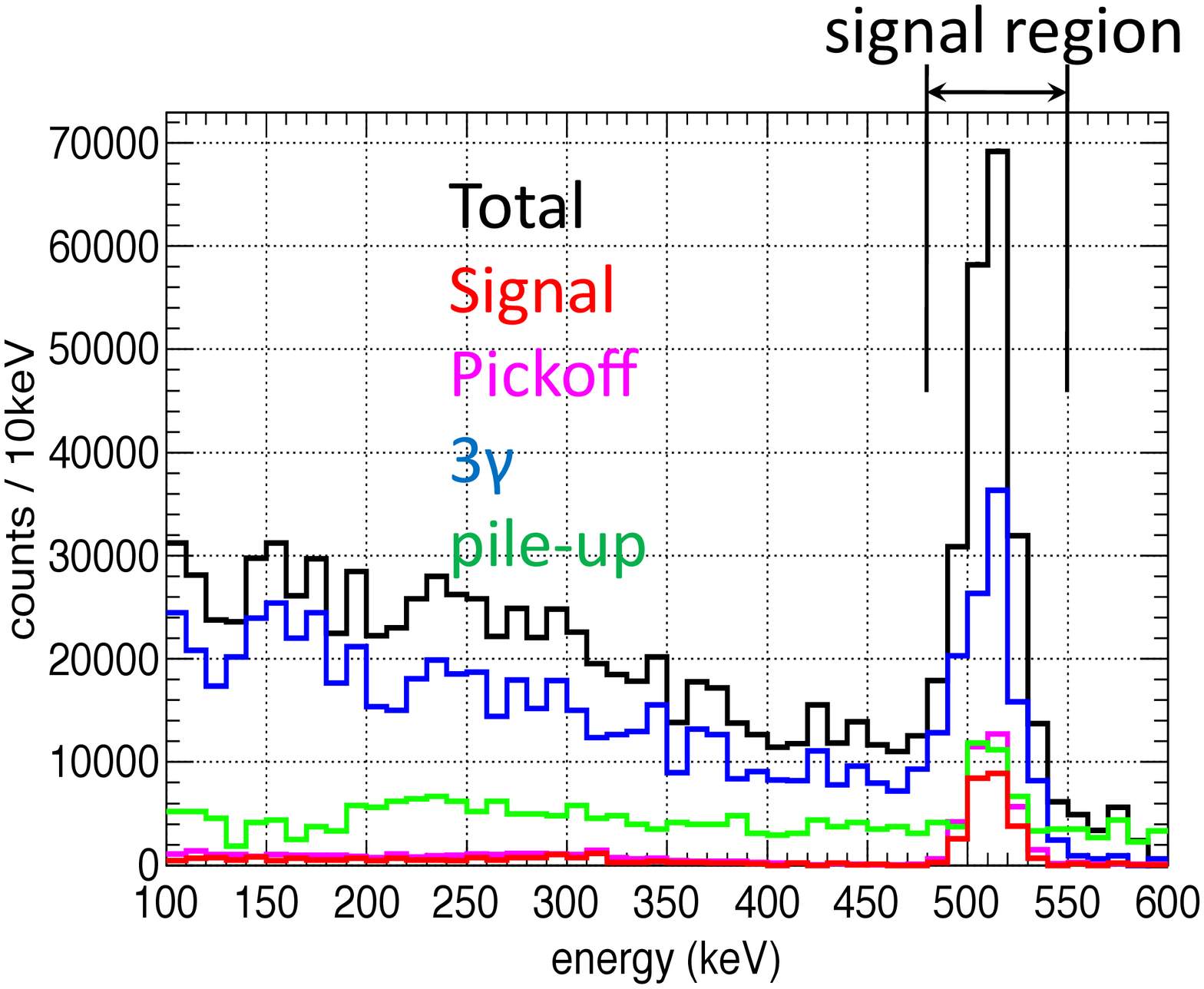}
\label{fig:signal}
\end{center}
\end{minipage}
&
\begin{minipage}{0.7\hsize}
\begin{center}
\begin{tabular}{cc}
\hline
process			&	rate (mHz)\\
\hline
signal			&	63	\\
pickoff			& 	93	\\
3$\gamma$		& 	300	\\
pileup			& 	100	\\
total (ON)	& 	560	\\
total (OFF)	& 	530	\\
\hline
\end{tabular}
\end{center}
\end{minipage}\\
Figure 13: Signal estimation&
\end{tabular}

\section{Summary}

There is a large discrepancy between theory and experiment in Ps-HFS value.
We suspect some common systematic errors in previous experiments.
The prototype experiments without RF system was already performed.
It was a complementary method against previous experiments,
and the accuracy of 200~ppm was obtained.
We are now tackling the direct measurement without any magnetic fields.
It is the first trial for sub-THz spectroscopy with $M1$ transition.
We have developed a high power 203~GHz radiation source called gyrotron, mode converter and Fabry-P\'erot cavity.
Monte Carlo simulation of the detection system shows that the observation of Ps-HFS is feasible.
We are now taking data of a test experiment for the direct transition from $o$-Ps to $p$-Ps.
The signal is expected to be observed in a month.
\\
\\
These experiments are collaborated with 
Y.~Sasaki, A.~Ishida, T.~Yamazaki, T.~Suehara, 
T.~Namba, S.~Asai, T.~kobayashi, 
H.~Saito, 
M.~Yoshida, K.~Tanaka, M.~Ikeno, A.~Yamamoto,
T.~Idehara, I.~Ogawa, Y.~rushizaki and S.~Sabchevski.


\end{document}